\documentclass[twocolumn,nofootinbib,prl,aps]{revtex4}
\usepackage{graphicx}

\begin{document}
\title{Conditional probabilities with Dirac observables and the problem of
time in quantum gravity}
\author{Rodolfo Gambini$^{1}$, 
Rafael A. Porto$^{2}$,
Jorge Pullin$^{3}$
and Sebasti\'an Torterolo$^{1}$
}
\affiliation {1. Instituto de F\'{\i}sica, Facultad de Ciencias,
Igu\'a 4225, esq. Mataojo, Montevideo, Uruguay. \\ 
2. Department of Physics, University of California, Santa Barbara, CA 93106\\
3. Department of Physics and Astronomy, Louisiana State University,
Baton Rouge, LA 70803-4001}
\date{September 24th 2008}

\begin{abstract}
  We combine the ``evolving constants'' approach to the construction
  of observables in canonical quantum gravity with the Page--Wootters
  formulation of quantum mechanics with a relational time for
  generally covariant systems. This overcomes the objections levied by
  Kucha\v{r} against the latter formalism. The construction is
  formulated entirely in terms of Dirac observables, avoiding in all
  cases the physical observation of quantities that do not belong in
  the physical Hilbert space.  We work out explicitly the example of
  the parameterized particle, including the calculation of the
  propagator.  The resulting theory also predicts a fundamental
  mechanism of decoherence.
\end{abstract}
\maketitle

In generally covariant systems, like general relativity, when one
works out the canonical formulation the Hamiltonian is a constraint,
i.e. it vanishes identically. This implies that the parameter that
usually plays the role of time in canonical formulations is not
adequate to describe the dynamics of the system. This constitutes one
of the aspects of the ``problem of time'' for generally covariant
systems (see Kucha\v{r} \cite{kuchar} for a good review).  Page and
Wootters proposed an approach to deal with this issue
\cite{PaWo}. The proposal consists in building a quantum theory of the
system of interest by promoting all variables of the system to quantum
operators and then choosing one of the variables to be a ``clock'' and
computing conditional probabilities for the other variables to take
certain values when the ``clock'' takes a given value.  This proposal
ran into technical difficulties when applied in detail to constrained
systems, as emphasized by Kucha\v{r} \cite{kuchar}. Basically, the
problem consists in what to choose as the variables to be promoted to
operators, in particular which one to choose as a ``clock''. In a
constrained system the physically observable variables are those that
have vanishing Poisson brackets with the constraints (this implies
they are invariant under the symmetries of the theory, they are
``gauge invariant''). However, if one of the constraints is the
Hamiltonian, then quantities that have vanishing Poisson brackets with
it do not evolve and therefore are poor candidates for being
clocks. Page and Wootters tried to circumvent this by considering
``kinematical'' variables (that do not have vanishing Poisson brackets
with the constraints and therefore appear ``to evolve'').  
But this brings about other problems. Such variables can
be promoted to quantum operators acting on the space of wavefunctions
that are not necessarily annihilated by the constraints (``kinematical
Hilbert space'').  Within such space, the states that are annihilated
by the constraints are usually distributional (at least in simple
examples that can be worked out explicitly). Therefore they may not
admit a probabilistic interpretation.  Kucha\v{r} showed by analyzing
the example of a parameterized particle, that these issues had as a
consequence the construction of propagators that ``do not propagate''
and therefore the resulting quantum theory is not realistic.

Here we would like to revisit the Page--Wootters construction but
using a different set of physical quantities. 
%In \cite{njp} (see also \cite{obregon}) we developed such construction
%within the scheme of consistent discretizations introduced in
%\cite{gapu}, where the conceptual problems of the continuum case do
%not appear and a well defined construction emerges. Even though the
%proposal of \cite{njp, gapu} is very appealing, here we will explore
%yet a different approach,
The quantities we will choose are relational Dirac observables such as
the ``evolving constants'' introduced in \cite{rovelli} (an idea that
goes back to DeWitt, Bergmann and Einstein).  The proposal can be
summarized as follows.  In a totally constrained theory, the values of
fields are not physically observable.  On the other hand, if one
chooses a one-parameter family of observables such that their value
coincides with the value of a dynamical variable when the parameter
takes the value of another dynamical variable, which one uses to
characterize the evolution, such observables can be used in the
Page--Wootters construction. They have the advantage that there is a
sense in which they ``evolve''.  That is, unlike the proposal of
Rovelli, we will not consider the ``parameter'' to be the {\it
physical} time, but we will use it to make sense of the conditional
probabilities that arise in the Page--Wootters formulation when one
introduces a real quantum clock. In fact at the end of the day the
parameter drops out from the formulation, and one integrates over all
possible values of it (if one has more than one constraint one needs
to introduce more parameters).  Therefore, one does not need to
observe any dynamical variable that is not quantum mechanical or is
not a Dirac observable.  We will show in an example that this
construction can be carried out in detail.

Let us start by defining the evolving constants in a classical
theory. Following Rovelli we consider a totally constrained system
with a phase space $q^i,p^i$. We now pick a parameter we call
$t$. We are interested in defining a one-parameter family of Dirac
observables that reproduces the value of one of the dynamical
variables, for instance $q^i$, when another variable playing the
role of a clock takes the value $t$. For concreteness, we can
choose $q^1$ to play the role of a clock. We denote the one
parameter family of Dirac observables $Q^i(t)=Q^i(t,q^n,p^n)$. 
These have vanishing Poisson
bracket with the Hamiltonian constraint,
$\{Q^i(t,q^n,p^n),C(q^n,p^n)\}=0$. They are also such that if one
evaluates $  Q^i(t,t,q^2\ldots q^n,p^n)\equiv q^i$. (We refer to the
observables as $Q^i$ for simplicity, they can include momenta
as well, but they must 
have vanishing Poisson bracket with the clock variable, an 
assumption that may be relaxed with further elaboration).

%From the definition one gets an evolution equation for the observables,
%\begin{equation}\label{evo}
%  {\partial Q^i(t)\over \partial t} \left\{q^1,C\right\} = 
%\left\{q^i,C\right\}.
%\end{equation}

 We then proceed to quantize the variables of the problem. Namely, we
 promote all $Q^i(t)$'s and $P^i(t)$'s to quantum self-adjoint
 operators acting on a Hilbert space of wavefunction $\Psi(q^i)$ that
 are annihilated by the constraints. The variable $t$ will remain
 classical. In realistic situations, like general relativity, this is
 convenient since usual choices of ``time'' are given by global
 variables that are not easily associated with a quantum operator to
 begin with. The restriction to self adjoint operators limits
 importantly the choices of possible parameters $t$, as was discussed
 in \cite{gapo}. In particular if one does not insist on
 self-adjointness one runs into problems in the definition of
 conditional probabilities (related to the ``false tracks'' discussed
 in \cite{hemopero}). If the quantization can be accomplished, then
 one can introduce a basis of eigenstates (parameterized by $t$) of
 the evolving constants and introduce projectors that materialize the
 physical properties associated to each of the evolving constants.
%One has therefore constructed a Heisenberg
% representation. One could attempt to construct a Schr\"odinger
% representation, but generically it may not be possible to introduce a
% Schr\"odinger equation, nor will the evolution of the states be given
% by a unitary transformation. However, there are many examples where
% translations in the $t$ parameter are given by a unitary
% transformation \cite{gapo}. Notice that the self-adjoint nature of
% the $\hat{Q}^i(t)$'s is enough to guarantee conservation of
% probabilities \cite{gapo}. Let us stress that unitarity is not needed
% to ensure it.
%The
%classical evolution equation (\ref{evo}) translates into a quantum
%evolution equation
%\begin{equation}
%  \label{eq:evoq}
%  {\partial \hat{Q}^i(t) \over \partial t} = {1 \over  \left\{q^1,C\right\}} 
%\left[\hat{q}^i,\hat{C}\right]
%\end{equation}
%which in turn can be used to create an evolution equation for the quantum
%states in the Schr\"odinger picture,
%\begin{equation}
%  \label{eq:schro}
%  i\hbar {\partial \Psi(q^i,t) \over \partial t} = 
%\end{equation}
%The quantization we just outlined would be a satisfactory quantum
%theory in regimes in which one can assume the existence of a classical
%clock and identify it with the parameter $t$. 

Having quantized the evolving constants, we will choose one of the
variables $Q^i(t)$ to be a quantum clock, and we will call it $T(t)$.
We  then introduce the conditional probability as in \cite{njp,obregon} 
\begin{equation}\label{probability}
 P{\scriptstyle \left(Q^i=Q^i_0 | T=T_0\right)} \equiv 
\lim_{\tau\to\infty} 
{\int_{-\tau}^{\tau} dt
{\rm Tr}\left( P_{Q^i_0}(t) P_{T_0}(t) \rho P_{T_0}(t)\right) \over
\int_{-\tau}^{\tau} dt
{\rm Tr}\left(  P_{T_0}(t) \rho \right)},\nonumber
\end{equation}
where $P_{Q^i_0}(t)$ is the projector on the eigenspace
associated with the eigenvalue $Q^i_0$ at time $t$ and
similarly for $P_{T_0}(t)$. These conditional probabilities are positive
and add to one.

By construction the conditional probability is a gauge invariant
quantity since the density matrix, $\rho$ in
the above expression, is assumed to be annihilated by the
constraints, e.g. $\hat{C} \rho =0$. Note that we are treating the variable
$t$ as an unobservable quantity and summing over all possible values
of it. This picture is much more satisfactory than the one that
emerges from considering evolving constants alone without the
conditional probability interpretation, since in that picture one
assumes that a quantum variable like $q^1$, takes a definite classical
value. This would not usually be the case since $q^1$ has a
non-vanishing Poisson bracket with the constraint and on the
constraint surface we expect $q^1$ to have infinite uncertainty.
Returning to the above expression, it should be noted that the
improper limits of integration may cause problems at the time of
computing the conditional probabilities. This can be controlled by
simply considering integrals in temporal domains that are much larger
than the region of physical interest.

It is worthwhile expanding on the meaning of the probabilities
(\ref{probability}) since there has been some confusion in the
literature \cite{hu}. One may interpret that the numerator of
(\ref{probability}) is the sum of joint probabilities of $O$ and $T$
for all values of $t$. This would be incorrect since the events in
different $t$'s are not mutually exclusive. The probability
(\ref{probability}) corresponds to a physically measurable quantity,
and that such quantity is actually the only thing one can expect to
measure in systems where one does not have direct access to the
``ideal'' time $t$. The experimental setup we have in mind is to
consider an ensemble of non-interacting systems with two quantum
variables each to be measured, $O$ and $T$. Each system is equipped
with a recording device that takes a single snapshot of $O$ and $T$ at
a random unknown value of the ``ideal'' time $t$. One takes a large
number of such systems, launches them all in the same quantum state,
``waits for a long time'', and concludes the experiment. The
recordings taken by the devices are then collected and analyzed all
together. One computes how many times $n(T_j,O_j)$ each reading with a
given value $T=T_j,O=O_j$ occurs (to simplify things, for the moment let 
us assume $T,O$ have discrete spectra; for continuous spectra one
would have to consider values in a small finite interval of the value
of interest). If one takes each of those values $n(T_j,O_j)$ and
divides them by the number of systems in the ensemble, one obtains, in
the limit of infinite systems, a joint probability $P(O_j,T_j)$ that
is represented by the above expression.

We can then write the conditional probabilities that yield the correlation
functions (propagators), namely the probability that having the system
been observed at $Q^i_1$ at time $T_1$ and 
it will be observed at $Q^i_2$ at time $T_2$,
\begin{eqnarray}
&&  P\left(Q^i_2 | T_2,Q^i_1,T_1,\rho\right) \equiv \\
&&\lim_{\tau\to\infty}
{\int_{-\tau}^{\tau}dt\int_{-\tau}^{\tau}dt'
{\rm Tr}\left( P_{Q^i_2,T_2}(t)P_{Q^i_1,T_1}(t')  \rho 
P_{Q^i_1,T_1}(t')\right) \over
\int_{-\tau}^{\tau}dt\int_{-\tau}^{\tau}dt'
{\rm Tr}\left(  
P_{T_2}(t)P_{Q^i_1,T_1}(t')  \rho 
P_{Q^i_1,T_1}(t')
\right)}.\nonumber
\end{eqnarray}
%This expression yields the propagator for the system to move from
%$Q^i_1,T_1$ to $Q^i_2,T_2$. 
Implicit is the use of a reduction postulate after the measurement
of $Q^i_1,T_1$ (see \cite{njp}).
Notice that in particular no assumption
about the relative ordering of the unobservable variables $t$ and $t'$
is needed.  We will show that it yields the correct propagator in an
example.

Up to now recovering the correct propagator has been problematic
in the conditional probability approach. For instance,
Kucha\v{r} \cite{kuchar} computed a similar expression
using the original Page and Wootters \cite{PaWo} prescription (in that
case, however there is no $t$ or $t'$ and no integrals over them)
where the quantities $Q^i$ were kinematical operators that did not
commute with the constraint and showed that one
obtained an incorrect propagator. Essentially, the system did not
move, the propagator being proportional to a Dirac delta function, e.g.
$\delta(Q^i_2-Q^i_1)\delta(T_2-T_1)$. Page \cite{Pa} has responded
to this criticism by claiming that in the conditional probability framework
one cannot compute two time probabilities.
We believe that the  framework can indeed accommodate such
probabilities and therefore becomes more powerful when formulated
in terms of evolving constants and indeed {\em  yields the correct
propagators.}

%It is convenient at this point to make some reasonable assumptions
%about the clock and the system. In particular we assume that they do
%not interact with each other so the quantum state of the joint
%clock-system under study can be represented as a tensor product 
%$\rho=\rho_{\rm clock}\otimes \rho_{\rm system}$ and that they in time
%$t$ via a unitary evolution that is also a tensor product
%$U=U_{\rm clock}\otimes U_{\rm system}$. One can then rearrange
%the probabilities in (\ref{probability}) as,
%\begin{equation}\label{probability}
%  P\left(Q^i=Q^i_0 | T=T_0\right) \equiv {\int_{-\infty}^{\infty} dt
%{\rm Tr}\left( P_{Q^i_0}(t) \rho \right) {\cal P}_t(T_0)\over
%{\rm Tr}\left( \rho \right)}.
%\end{equation}
%where $\rho$ is what we called above $\rho_{\rm system}$ and 
%${\cal P}_t(T_0)={\rm Tr}\left(P_{T_0}(t)\rho_{\rm clock}\right)$. 

The example we will consider is a simple model of two non-interacting
non-relativistic free particles in one spatial dimension that has been
``parameterized'', that is, Newtonian time is introduced as a
canonical variable conjugate to the energy.  The reader may question
how relevant these simplified examples are to the issue of interest,
namely the problem of time in quantum gravity.  To quote Kucha\v{r}
\cite{kuchar} {\em ``The nature of the conditional probability
interpretation is so clear from these examples that it is hardly
necessary to spell out how the formalism looks in quantum gravity''}.
The reader will confirm this point of view while seeing how one gets
the result for the propagator virtually without using any special
features of the model in question. In particular, although the model
does have a naturally defined time variable, we only use it to
construct easily the evolving constants. The latter are known to exist
in many examples (e.g. \cite{gapo}) where there is no natural
decomposition of the constraint into the ``$p^0+H$'' form.

The system has three configuration variables $q_0, q_1,q_2$ and the
corresponding canonical momenta $p_0,p_1,p_2$. There is a constraint
$\phi=p_0+p^2_1/(2m_1) +p^2_2/(2m_2)$. The gauge invariant quantities, which have
vanishing Poisson brackets with the constraint, are $Q_1=q_1-p_1
q_0/m_1$ and $Q_2=q_2-p_2 q_0/m_2$ and $p_1$ and $p_2$.  These
Dirac observables represent the initial position and momenta of the
particles. We then define evolving constants $X_1(t)=Q_1+p_1 t/m_1$
and $X_2(t)=Q_2+p_2 t/m_2$. We can check that they have vanishing
Poisson bracket with the constraint and that when $t=q_0$ then
$X_1(t=q_0)=q_1$ and $X_2(t=q_0)=q_2$. The quantization of the model
is immediate \cite{gapo}. The states that are annihilated by the quantum version
of the constraints are given by $\psi(p_1,p_2)$ times a prefactor
$\delta(p_0+p^2_1/(2m_1)+p^2_2/(2m_2))$ and the Hilbert space is
that of square integrable functions $\psi(p_1,p_2)$, or equivalently in Fourier space by functions $\tilde \psi(q_1,q_2)$. In this Hilbert
space the evolving constants are well defined operators. Their
common eigenstates are of the product form,
\begin{eqnarray}
&&\psi_{x_1,x_2;t}=\langle p_1,p_2\vert x_1,x_2;t\rangle\\&=&\frac{1}{2\pi} \exp\left(
-i\left[p_1x_1+p_2 x_2-t\left(\frac{p^2_1}{2 m_1}+\frac{p^2_2}{2 m_2}\right)\right]\right),\nonumber
\end{eqnarray}
with eigenvalues $x_1,x_2$ for $\hat{X}_1,\hat{X}_2$, at some  value of $t$. 
With these we can construct the projectors that appear in the conditional
probability, $P_{x_1}(t) =\int_{x_{1-}}^{x_{1+}} dz_1
\int_{-\infty}^{\infty}  dz_2 \vert z_1,z_2;t\rangle\langle z_1,z_2;t\vert$
and similarly for $P_{x_2}(t)$. 
The limits of integration $x_{i\pm}$ correspond to
$x_{i}\pm \Delta x_i/2$ where $\Delta x_i$ is introduced since if one is
dealing with variables that have continuum spectrum, one cannot ask
for ``the probability that $q_i$ is a given value $x_i$, but rather within an
interval of width $\Delta x_i$ centered at such value.
Let us consider a physical state given by a Gaussian for both
variables centered at two distant phase-space 
points ${x}^0_1$ and ${x}^0_2$ and ${p}^0_1$ and ${p}^0_2$, 
e.g. $\rho_0=\vert\psi_0\rangle\langle \psi_0\vert$, with
$\tilde \psi_0(q_1,q_2)= \prod_{j=1}^2
\exp\left(-(q_j-{x}^0_j)^2/\Delta_j^2
+i{p}^0_j q_j\right)$. We can then compute the quantity in the
numerator of (\ref{probability}), (notice that the denominator is just
given by the numerator integrated in $Q^i_2$ from $-\infty$ to $\infty$, so
for brevity we only show explicit calculations for the numerator),
\begin{eqnarray}
&&{\rm Num}\left(
P\left(x'_2 | x'_1,x_2,x_1\right),\rho_0\right)=\int_0^\tau dt' dt \\
&&\times\int_{x_{2-}}^{x_{2+}}dy_2{dz}_2
\langle x'_2, t'\vert y_2, t\rangle\psi^0_{y_2,t}\left(\psi^0_{z_2,t}\right)^*
\langle z_2,t\vert x'_2,t'\rangle\nonumber\\
&&\times\int_{x_{1-}}^{x_{1+}}dy_1dz_1
\langle x'_1, t'\vert y_1, t\rangle\psi^0_{y_1,t}\left(\psi^0_{z_1,t}\right)^*
\langle z_1,t\vert x'_1,t'\rangle\nonumber
\end{eqnarray}
where $\psi^0_{w_i,t} \equiv \langle w_i,t | \psi^i_0\rangle$ for
$i=1,2$, and we have used the fact that the density matrix for this
model is of direct product form, namely $|\psi_0\rangle
=|\psi^1_0,\psi^2_0 \rangle$. Usually one would like to consider
systems with this property which implies that the system under study
and the clock do not interact (we are choosing $x_1$ as the clock
variable).  The interval $\Delta x_1$ must be taken much larger than
$\Delta_1$ the width of the Gaussian in the state in order for the
measurement of the clock variable not to ``destroy the state of the
clock''.  A measurement with more precision implies a faster loss of
the (desired) classicality of the clock.  In the case of $x_2$ we
assume we are studying a microscopic variable ($m_2 \ll m_1$),
i.e. that is behaving quantum mechanically, therefore we may and will
assume $\Delta x_2$ much smaller than the width of the Gaussian
$\Delta_2$ to simplify the calculation of the integrals by
substituting mean values. Carrying out the integrations explicitly,
\begin{eqnarray}
&&{\rm Num}\left(
P\left(x'_2 | x'_1,x_2,x_1\right),\rho_0\right)\sim\int_0^\tau dt' dt 
\vert\langle x'_2,t'\vert x_2,t\rangle \vert^2\nonumber\\
&& \times\vert \psi^0_{x_2,t}\vert^2
\Delta x^2_2 \Theta_{\Delta x_1}\left(\bar{x}^0_1-\frac{p_1}{m_1} t -x_1\right)
\vert\psi^0_{x'_1,t'}\vert^2
\end{eqnarray}
where $\Theta$ is a rectangular function that is unity in the interval of 
width $\Delta x_1$ around its argument and zero otherwise. 
We have assumed that evolution times are small such that the
value of $\Delta x_1$ does not change significantly.
We have approximated the integrals in $y_1$ and $z_1$ by 
integrals from $-\infty$ to $\infty$ since the Gaussian has a smaller
support than the region of integration. The $\Theta$ function arises
since the approximation is good only if the peak of the 
Gaussian is within the integration region, otherwise the integral is 
close to zero. Putting together numerator
and denominator we get,
\begin{eqnarray}
&&P\left(x'_2 | x'_1,x_2,x_1,\rho_0\right)\sim\\ \label{propagator}
&&\lim_{\tau\to\infty}
\int_0^\tau dt' \vert\langle x'_2,t'\vert x_2,t(x_1)\rangle\vert^2 
{\cal P}_{x'_1}(t') 
\Delta x_2\nonumber 
\end{eqnarray}
where $t(x_1)$ is the central 
value of $t$ determined by the $\Theta$ function
and ${\cal P}_{x'_1}(t') \equiv {\rm Tr}
\left(P_{x'_1}(t') \rho_0\right)/\int_{-\infty}^\infty dt {\rm Tr} 
\left(P_{x'_1}(t) \rho_0\right)
$ can be interpreted as the probability that the external
(unobservable) time $q^0$ is $t'$ when the variable we take as clock
reads $x'_1$. This would be controlled by the position of the peak and
width of the Gaussian in the quantum state we chose.  If instead of a
Gaussian one had a Dirac delta, then we would recover the correct
ordinary non-relativistic propagator, $P\left(x'_2 |
x'_1,x_2,x_1,\rho_0\right)\sim\langle x'_2,t'(x'_1)\vert
x_2,t(x_1)\rangle \Delta_2,$ where $t(x_1)$ is determined by our
choice of initial state to approximate the ordinary non-relativistic
time corresponding to the position $x_1$. The resulting expression is
an approximation to the integral in $x_2$ of the ordinary propagator,
therefore the factor $\Delta x_2$. As is expected in relational
treatments, one only obtains the traditional propagator at leading
order. The use of real clocks leads to loss of quantum coherence, as
is well known
\cite{obregon} and therefore to corrections to the ordinary
propagator. Notice that up to now discussions of loss of coherence due
to real clocks did not involve the presence of constraints, since they
were framed for the gravitational case in the context of uniform and
consistent discretizations \cite{njp}, where constraints are
eliminated. Here we confirm the presence of these effects in totally
constrained systems.

Let us sketch how the above proposal could be implemented in the case
of general relativity. We consider the theory in vacuum coupled to a
clock. We characterize the clock by its worldline $X^\mu(\tau)$ and
$T(\tau)$ its proper time. The action is the Einstein--Hilbert action
for general relativity plus a term for the clock of the form $S=-m
\int d\tau \sqrt{-\dot{X}^\mu \dot{X}^\nu
g_{\mu\nu}(X(\tau))-\dot{T}^2}$ where the dots mean total derivative
with respect to the parameter $\tau$, and $m$ is the mass of the
clock. The equations of motion state that $X^\mu$ is a geodesic of the
metric $g_{\mu\nu}$ and an equation stating that $T$ is proportional
to the proper time. As usual, we are assuming that the clock is a
probe and therefore ignore back reaction. Classically this is
certainly a good approximation.  In this system we have only
introduced a clock, not a complete coordinate system, one can ask only
certain relational questions. For instance what is the value of a
geometric scalar (e.g. the Kretschmann invariant) $K$ at the
space-time position of the clock when the clock measures a given value
of time $T$. To complete our proposal one needs to find evolving
constants $S$, functions of the metric and its first derivatives,
parameterized with {\em four} real parameters $x^\mu$ such that when
they equal certain combinations of the metric and its derivatives the
evolving constants reproduce the geometric quantity $S$ we want to
measure. The explicit construction of these quantities in general
relativity can be onerous, but progress can be done by perturbative
techniques, for example (see \cite{dittrich}).  One can then define
the relational probabilities that the geometric quantity of interest
take the value $S_0$ when the clock measures time $T_0$,
\begin{eqnarray}
P(S_0\vert T_0)\!\!=\!\!\frac
{\int d^4x  {\rm Tr}(\sqrt{-g} \rho)
{\rm Tr}\left(P_{S_0}(x) P_{T_0}(x) \rho P_{T_0}(x)\right)}
{\int d^4x  {\rm Tr}(\sqrt{-g} \rho)
{\rm Tr}\left(P_{T_0}(x)\rho\right)}.\nonumber
\end{eqnarray}

Defining a propagator needs more work, namely
setting up a full coordinate system (i.e. introducing rulers in
addition to clocks or considering a cloud of clocks as in \cite{rovelli2}).  The calculational
complexity would be important but modern loop quantum gravity
techniques may allow a proper calculation. The expressions of the
conditional probabilities in a situation like general relativity will
not only include loss of coherence in time but also spatially, as has
been analyzed in field theory in \cite{spatial}.

Summarizing, we have shown that one can formulate a
completely relational picture of evolution in generally
covariant systems framed entirely in the physical space
of states and that yields the correct propagators in 
model systems and opens the possibility of 
assigning probabilities to  histories and characterizing
consistently the dynamics of quantum general relativity. The resulting theory
also predicts a fundamental mechanism of decoherence similarly as the one originally discussed 
in \cite{njp}.

We wish to thank Don Marolf for detailed comments.
This work was supported in part by grant NSF-PHY-0650715,
funds of the Hearne Institute for Theoretical Physics, FQXi, RPFI-06-18, CCT-LSU, 
the University of California, Pedeciba and PDT (Proyecto 63/076)  (Uruguay).

\end{document}